\documentclass[seceq,supplement]{ptptex} \usepackage{graphicx}
\usepackage{wrapfig,wrapft}

\title{Cluster Monte Carlo algorithms for diluted spin glasses} 
\author{Thomas \textsc{J\"org}}
\inst{Dipartimento di Fisica, INFM and SMC, Universit\`{a} di Roma "La Sapienza",\\ 
  P.~A.~Moro 2, 00185 Roma, Italy} 
\recdate{\today} 
\abst{Recently a cluster Monte Carlo algorithm has been used very 
  successfully in the two-dimensional Edwards-Anderson (EA) model. We 
  show that this algorithm and a variant thereof can also be used successfully
  in models with a non-zero spin glass transition temperature. The
  application of such algorithms to the site-diluted EA model in three
  dimensions is discussed and the efficiency of the two algorithms is
  compared among each other and to parallel tempering. Finally, we 
  give evidence for a spin glass transition in the three-dimensional 
  site-diluted EA model with Gaussian couplings at a site occupation of $p = 62.5\%$.}

\begin{document}

\maketitle

\section{Introduction}

The determination of equilibrium properties of finite dimensional spin glass models
relies heavily on Monte Carlo (MC) simulations. However, these simulations have
shown to be a very difficult task due to the strong critical slowing down
at the phase transition and the existence of big free energy
barriers in the low temperature phase. There are few algorithms that
sample the relevant regions of the phase space in spin glass models
efficiently enough to allow for the simulation of systems with sizes
around $16^3$ to $20^3$ spins with reasonable statistics.

The use of cluster algorithms which revolutionized the simulations in
ordinary spin models \cite{SwendsenWang87,Wolff89} hasn't shown to be very effective in spin glass
models due to frustration and the lack of simple symmetries.
Recently, however, Houdayer introduced a cluster algorithm combined with the exchange MC
method (also known as parallel tempering (PT)) which has shown to be extremely
efficient in the two-dimensional Edwards-Anderson (EA) model.\cite{Houdayer01}\footnote{A slightly modified version of
the original Swendsen-Wang replica cluster algorithm has even shorter autocorrelation times 
than Houdayer's algorithm in the two-dimensional EA model \cite{Wang04}.}  The main
idea of this algorithm is already present in \citen{SwendsenWang86,RednerMachta98} and is to
simulate two or more replicas of the same system and to use their
mutual overlap to generate clusters.  The main drawback mentioned in
\citen{Houdayer01} is that the algorithm performs poorly for the
three-dimensional EA model because of a site percolation problem and therefore the method seemed to work
only for systems with a spin glass transition at zero temperature.

In this paper we show that Houdayer's algorithm 
and a related algorithm described below do also work efficiently in a
certain class of spin glass models with a finite spin glass transition
temperature. Here we will focus on the three-dimensional site-diluted
EA model given by the Hamiltonian\cite{EdwardsAnderson75}
\begin{equation}
  \label{eq:diluted_EA}
  H = - \sum_{<ij>} J_{ij} \epsilon_i \epsilon_j \sigma_i \sigma_j ,
\end{equation}
where the sum is over all the nearest neighbor sites of a simple cubic
three-dimensional lattice of length $L$, the $\sigma_i = \pm 1$ are Ising spins and
the $\epsilon_i$ are the site occupation variables, i.e.~$\epsilon_i =
0$ for an empty and $\epsilon_i = 1$ for an occupied site.  For the
couplings $J_{ij}$ we use a Gaussian distribution with mean zero and
standard deviation unity.
The fraction of occupied sites is denoted by
$p$.  At this point we would like to mention that there are other
diluted models for which such replica cluster (RC) algorithms work very
efficiently as e.g.~the Viana-Bray model\cite{VianaBray85} with fixed connectivity $z=3$
and also the link-diluted EA model in three dimensions\cite{Shapira94,Boettcher03}.

The site-diluted EA model in three dimensions is clearly a model of physical relevance\footnote{The EA model in the original work\cite{EdwardsAnderson75} is actually formulated as a site disordered model.}. 
One prototype of a real spin glass consists of a small fraction of randomly diluted magnetic moments dispersed 
in a non-magnetic metal and hence site disorder is an inherent property of such a system.
It is therefore desirable to check numerically whether the EA model is stable with respect
to site-dilution, as one might expect on general grounds\cite{Harris74}.
Note that this is in contrast to the expectations in the site-diluted Ising model, 
where site-dilution shows to be a relevant perturbation that changes the 
critical behavior of the system \cite{MartinMayor98}. In the following we are 
going to discuss a variation of Houdayer's RC algorithm used in this study.
We show that it is very efficient in thermalizing the spin glass model defined through
Eq.~\eqref{eq:diluted_EA}, compare it to the original version of Houdayer's algorithm\cite{Houdayer01}
and to PT \cite{HukushimaNemoto96}.  Finally, we discuss possibilities and limits of 
this kind of RC algorithms and show results for the finite-size scaling of the correlation length.

\section{Algorithm and its limitations}
  
The idea of the algorithm is as follows. Two independent replicas of
the system are simulated simultaneously.  Hence each site of the
lattice is associated with two spins, one from each replica, and can
be in one of four spin states $(++)$, $(+-)$, $(-+)$, and $(-\,-)$. We
call the sites where the replicas disagree, $(+-)$ and $(-+)$, active
sites. In the case of a non-zero external magnetic field clusters of
active sites are constructed and flipped on both replicas. Such cluster flips
interchange $(+-)$ with $(-+)$ and vice versa.  In the case of no
external magnetic field also the sites where the replicas agree
(non-active sites) can be used to construct clusters and we will
comment on this possibility below. In order to ensure ergodicity of
the algorithm we update each replica independently by wrapping the
cluster moves within a PT as proposed by Houdayer \cite{Houdayer01}.

The clusters themselves are formed as described in
\citen{RednerMachta98}, i.e.~they are Wolff-type clusters grown on the
active sites of the lattice.  However, having a disordered Hamiltonian
we need to generalize the procedure slightly as follows. The clusters are grown on 
the active sites by adding links with probability $p_{add}$:
\begin{equation}
  \label{eq:link_addition}
  p_{add} =
  \begin{cases} 
    1 -\exp(4 \beta E_{ij})\quad&\text{if}\quad E_{ij} < 0 \\
    0\quad&\text{else}, \\
  \end{cases}
\end{equation}
where $E_{ij} = - J_{ij} \epsilon_i \epsilon_j \sigma_i \sigma_j$ is
the energy contribution of the given link and $\beta$ is the inverse
temperature. This is the only difference to Houdayer's algorithm, 
where the links between two active sites are always added to the cluster. 
The clusters we have defined through Eq.~\eqref{eq:link_addition} flip freely i.e.~the cluster move is always accepted. Note
that also more general clusters might be constructed along the lines
of \citen{Niedermayer88}, where the size of the clusters can be
controlled in a very general manner. In the absence of an external
field one may as well grow clusters on the non-active sites, as the
constraint to the active sites is just to ensure that the total magnetic
energy of the two replicas is conserved, as $(+-)$ goes to $(-+)$ and
vice versa. Without external field one may hence also do a full
Swendsen-Wang updating of all clusters, which in diluted systems may
be helpful to update clusters of all sizes\cite{MartinMayor98}.

The algorithm we proposed is a valid MC procedure for general Ising
spin systems. However, for frustrated interactions it will generally
not perform well when the site percolation threshold $p_{\rm SP}$ of the
underlying interaction graph is below $0.5$, because in this case only very small (or equivalently very large) clusters are formed. If, on the other hand, 
$p_{\rm SP} \geq 0.5$ clusters of all sizes can be formed in the disordered phase, allowing for a fast decorrelation of the configurations.

\section{Results and Conclusions}

\begin{figure}
  \begin{center}
    \includegraphics[width=5.9cm]{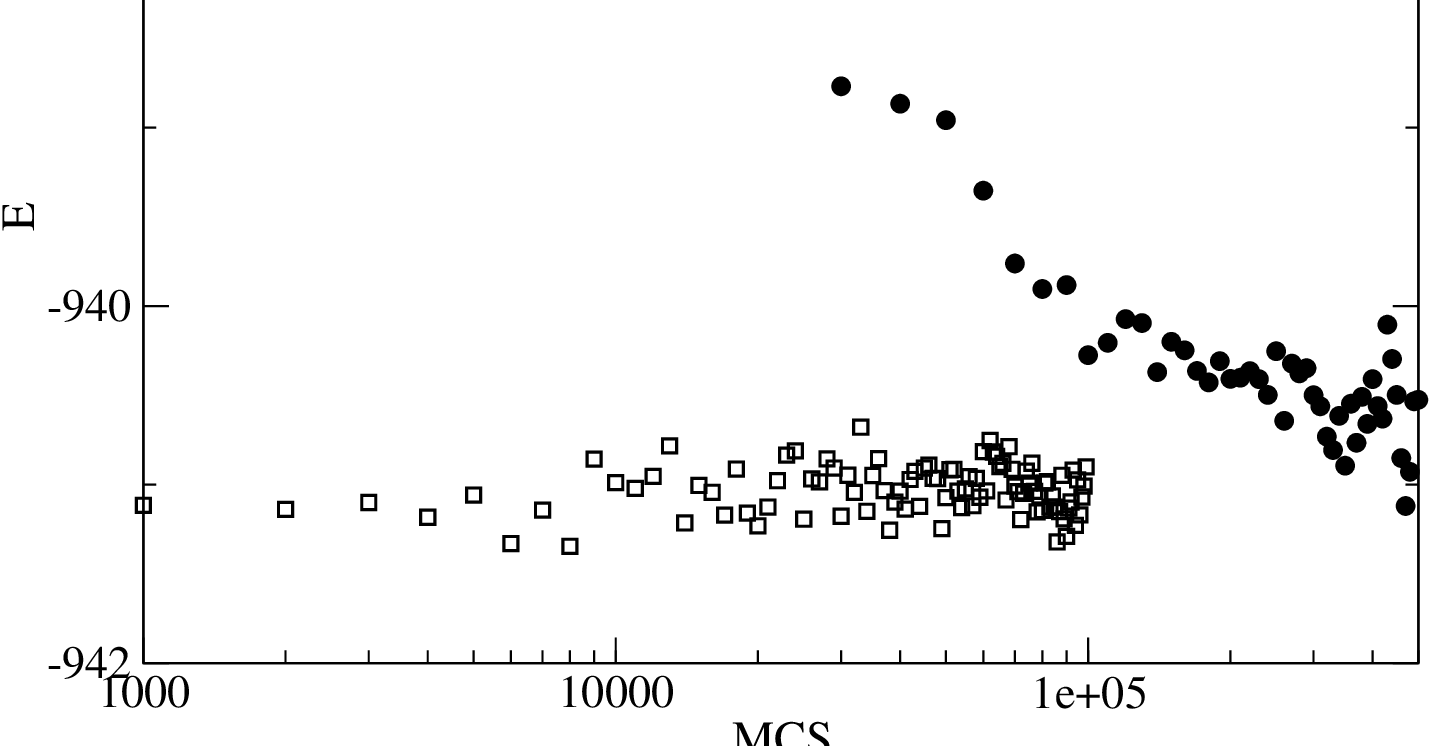}\hspace{1cm}
    \includegraphics[width=5.9cm]{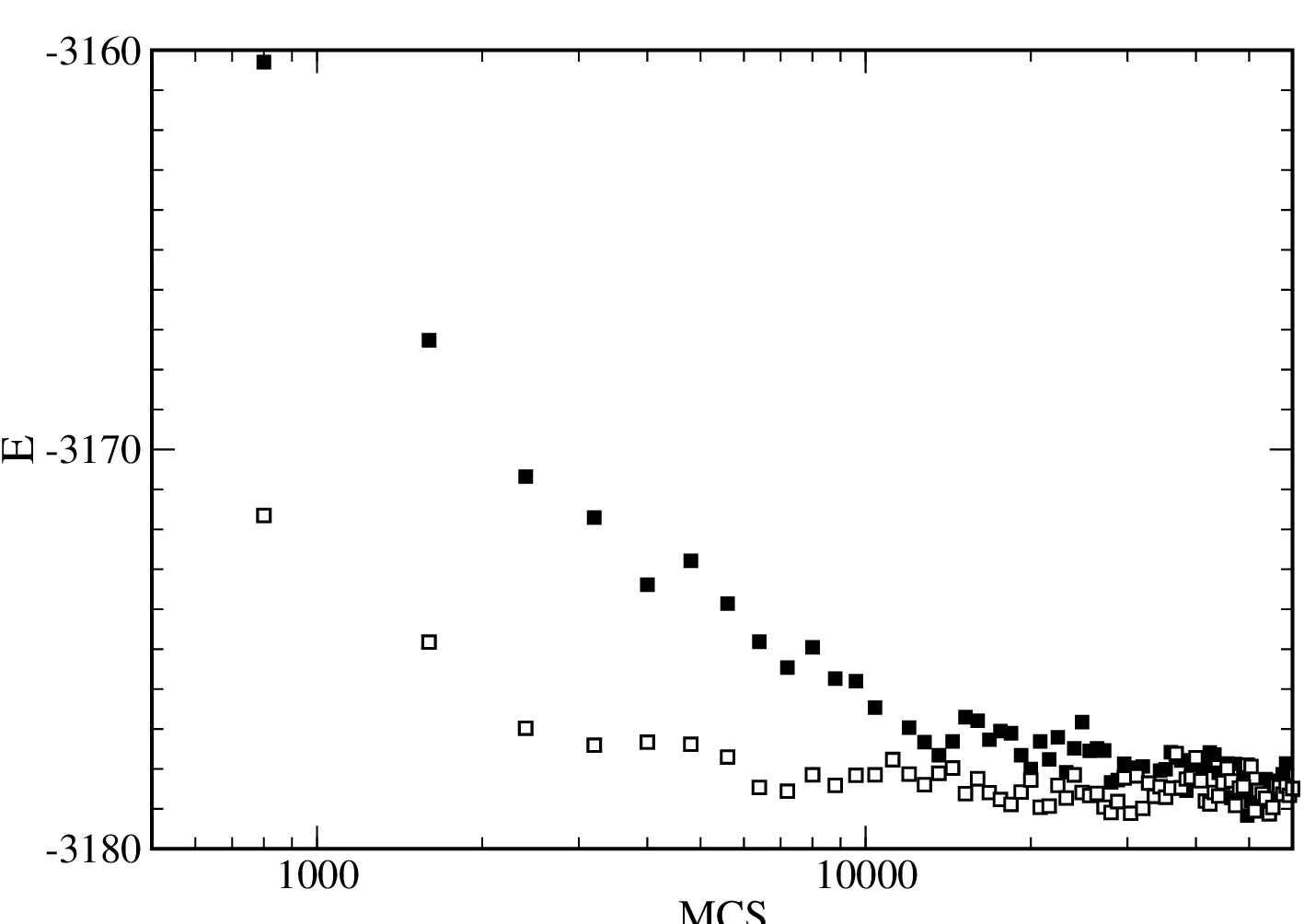}
  \caption{Equilibration of the energy. Left panel: Comparison of the RC algorithm with Wolff-type clusters (open squares) with PT (filled circles).
           Right panel: Comparison of the RC algorithm with Wolff-type clusters (open squares) with Houdayer's clusters (filled squares). Note that the 
           two comparisons are done on different system sizes.}
  \label{fig:therm_energy}
  \end{center}
\end{figure}

For the three-dimensional site-diluted EA model the RC algorithms are
performing clearly better in thermalizing a configuration than normal PT, 
as illustrated in the left panel of Fig.~\ref{fig:therm_energy} for a $12^3$ system with $p=0.5$ at roughly $0.6 T_c$.  
Although one MC step in a RC algorithm is more expensive than for PT, the
performance in absolute computer time is by far better for the RC algorithm.
The right panel of Fig.~\ref{fig:therm_energy} shows how the cluster algorithm
defined through Eq.~\eqref{eq:link_addition} compares to Houdayer's algorithm. The equilibration of the
energy on a $16^3$ system with $p=0.625$ at roughly $0.8 T_c$ is significantly faster for the Wolff-type 
clusters.
\vspace{0.3cm}

\begin{wrapfigure}{r}{6.0cm} 
  \includegraphics[width=5.9cm,height=5.9cm]{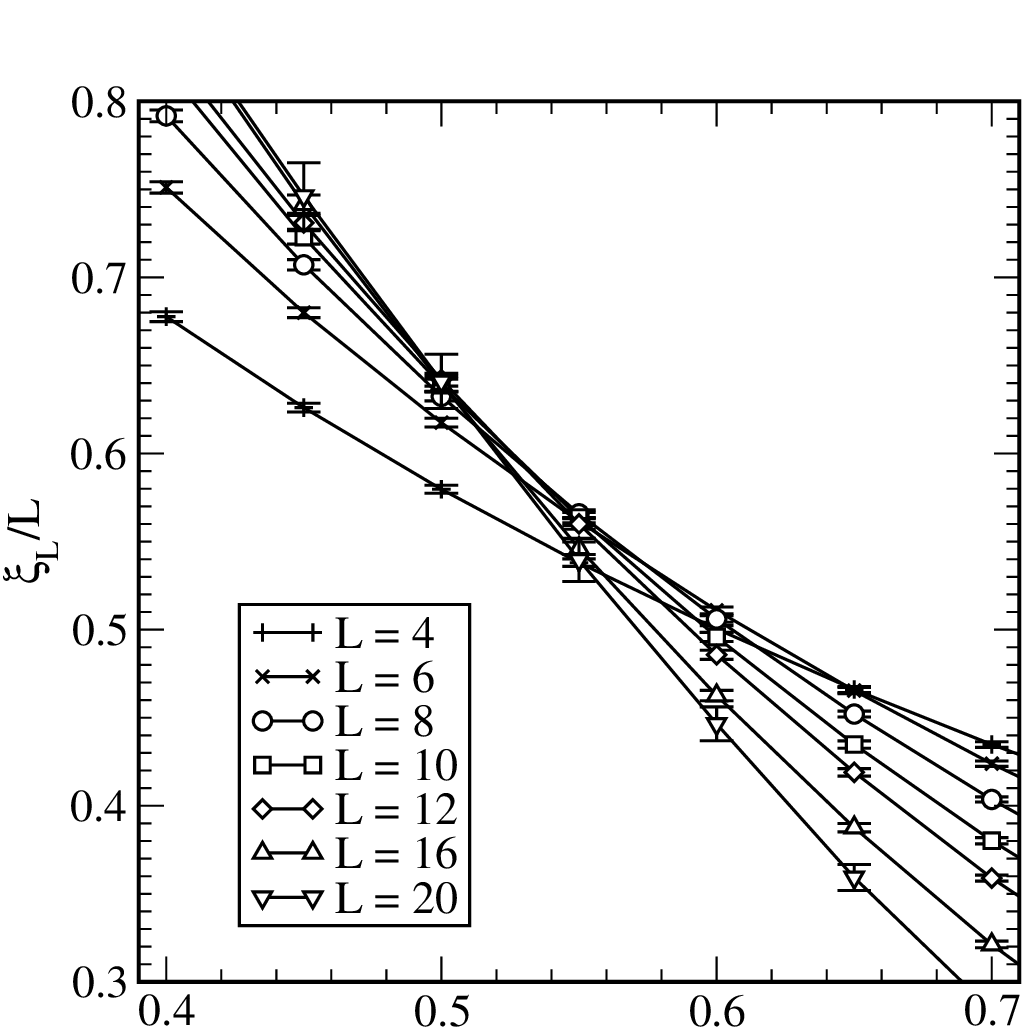}
  \caption{Data for the correlation length $\xi_{L}/L$ for p = 0.625.}
  \label{fig:corr_length}
\end{wrapfigure}

Finally, Fig.~\ref{fig:corr_length} shows the finite size behavior of the overlap
correlation length $\xi_L/L$, as defined in \citen{Palassini99}, for system sizes ranging from $4^3$ to
$20^3$ at a site dilution of $p = 0.625$.  For small systems the curves do not cross at a unique point due to
finite size effects. For larger systems, however,
the curves start to cross very close to each other, such that this data presents clear
evidence for a phase transition at $T_c \sim 0.51$. The value of $\xi_L/L$ at $T_c$ is in good agreement with 
the simulations of the undiluted three-dimensional $\pm J$\cite{Ballesteros00} and Gaussian\cite{Young04} EA model.
More details are given in \citen{Jorg04}.

RC algorithms work efficiently in some systems with a
non-zero spin glass temperature, typically in models with a low connectivity. 
This opens a new perspective to investigate spin glass models that have not been well accessible by
more traditional MC methods, like site- or bond-diluted EA
models in three dimensions.  Moreover, we have shown that using a Wolff-type
definition of the clusters can have a benefit with
respect to Houdayer's definition. A more precise determination of the critical properties of the three-dimensional 
site-diluted EA model will allow to give clearer results about its universality class. Finally, 
there remains the question whether the fact that efficient cluster algorithms for some very specific spin 
glass models exist is just an incident or not.

\section*{Acknowledgments}

It's pleasure to thank F.~Ricci-Tersenghi, F.~Krzakala and F.~Niedermayer for interesting discussions, the institute of
theoretical physics in Bern for the use of the workstations and the European Community's Human
Potential program under contract HPRN-CT-2002-00307 (DYGLAGEMEM) for financial support.

\end{document}